\documentstyle[preprint,aps,eqsecnum,epsf]{revtex}
\tightenlines
\begin{document}
\title{Coupling Constant and Quark Loop Expansion for 
Corrections to the Valence Approximation}
\author{W. Lee\cite{LANL} and D. Weingarten}
\address{IBM Research,
P.O. Box 218, Yorktown Heights, NY 10598}
\maketitle

\begin{abstract}

For full QCD vacuum expectation values we construct an expansion in
quark loop count and in powers of a coupling constant. The leading term
in this expansion is the valence (quenched) approximation vacuum
expectation value. Higher terms give corrections to the valence
approximation. A test of the expansion is presented for moderately heavy
quarks on a small lattice. We consider briefly an application of the
expansion to quarkonium-glueball mixing.

\end{abstract}


\narrowtext

\section{INTRODUCTION}\label{sect:intro}

The infinite volume, continuum limit of lattice QCD hadron
masses~\cite{Butler_mass,MILC,CPPACS} and meson decay
constants~\cite{Butler_decay,CPPACS} calculated in the valence
(quenched) approximation lie not far from experiment. Calculations using
the valence approximation, however, require significantly less computer
time than those using full QCD. Thus, in at least some cases, the
valence approximation can serve as a cheap, approximate substitute for
full QCD. For this purpose it would be useful to have some way to
determine quantitatively an estimate of the error arising from the
valence approximation short of a direct comparison of the valence
approximation with full QCD.

A possible method for finding the valence approximation's error is given
in Ref.~\cite{Sexton}. In the present article, we describe an alternative
form of the proposal in Ref.~\cite{Sexton} which we believe will
generally require less computer time.  The expansion we describe can be
applied to any choice of quark action but is given here only for Wilson
quarks.

In full QCD, virtual quark-antiquark pairs produced by a chromoelectric
field reduce the field's intensity by a factor which depends both on the
field's momentum and on its intensity.  In the valence approximation
this factor, analogous to a dielectric constant, is approximated by its
zero-field-momentum zero-field-intensity limit~\cite{DW82}.  Our
expression for the error in valence approximation vacuum expectation
values consists of an expansion in quark-loop count and in powers of a
coupling constant. The coupling constant expansion relies on ideas drawn
from mean-field-improved perturbation theory~\cite{Lepage}.  Each term
in the expansion requires as input the quantity $\Delta \beta$ given by
$(6 \eta^2/g^2) - (6/g^2)$, where $g$ is the gauge coupling constant of
full QCD and $\eta$ is the dielectric constant entering the valence
approximation.  We determine $\Delta \beta$ analytically from
mean-field-improved perturbation theory to second order in the coupling
constant. A related calculation of $\Delta \beta$ without mean-field
improvement is described in Ref.~\cite{Hasenfratz}.  The remaining work
of evaluating each term in the error expansion is done by a Monte Carlo
algorithm.

The sum of all terms in the error expansion, in principle, gives the
exact value of the valence approximation error for any choice of the
$\Delta \beta$.  In particular, the expansion remains correct
independent of the accuracy of the second order perturbative expression
for $\Delta \beta$.  For a bad choice of $\Delta \beta$, however,
valence approximation vacuum expectations will be far from their full
QCD values and the error expansion will predict an error correspondly
large.

We have tested our method so far only for vacuum polarization arising
from quarks with about 1.8 times the strange quark mass and only for a
collection of Wilson loop expectation values.  For these cases, our
method of estimating the valence approximation error is significantly
faster than direct comparison between the valence approximation and
full QCD. A test of the efficiency of our method for lighter quark
masses and other vacuum expectation values we hope to return to
elsewhere.

In addition to its use as an algorithm for finding valence approximation
errors, the expansion we describe provides a systematic way to keep
track of the quantities which need to be evaluated, by any method, to
determine quark loop corrections to valence approximation vacuum
expectation values.  Toward the end of the present article, we present a
brief, qualitative discussion of the valence approximation and
corrections to the valence approximation for mixing between the lightest
scalar glueball and scalar quarkonium states.  We will show that a
recent attempt~\cite{Boglione} to determine glueball-quarkonium mixing
misses two of the terms required for this calculation. As a consequence, 
we believe the calculation of Ref.~\cite{Boglione} is not correct.

In Section~\ref{sect:defs} we introduce definitions.  In
Section~\ref{sect:exp}, we construct an expansion for the valence
approximation error. In Section~\ref{sect:deltabeta} we discuss the weak
coupling calculation of the shift between the coupling constant in full
QCD and in the valence approximation.  In Section~\ref{sect:examp} we
describe a trial calculations using our expansion and error estimates.
In Section~\ref{sect:mix} we consider the valence approximation and its
corrections for glueball-quarkonium mixing.

\section{DEFINITIONS}\label{sect:defs}

For euclidean QCD on some finite lattice, let $u( x, y)$ be a guage
link with periodic boundary conditions, and let $M$ be the
coupling matrix for a single quark flavor, with antiperiodic boundary
conditions, defined by
\begin{eqnarray}
\label{defM}
M(x,y) & = & \delta_{x y} - \kappa \sum_{\mu} \delta_{x y - \hat{\mu}} 
(1 - \gamma_{\mu}) u(x,y)
- \kappa \sum_{\mu} \delta_{x y + \hat{\mu}} 
(1 + \gamma_{\mu}) u(x,y).
\end{eqnarray}
The vector $\hat{\mu}$ is a unit lattice vector in the $+\mu$ direction
and the $\gamma_{\mu}$ are $4 \times 4$ hermitian euclidean
gamma-matrices.

For $n_f$ degenerate flavors of quarks and any integrable function of
the gauge fields $G$, the vacuum expectation value found after
integrating out quark fields becomes
\begin{eqnarray}
\label{expG}
< G > & = & Z^{-1} \int d \nu \, G \, det( M)^{n_f} \, 
          exp( \frac{\beta}{6} P ), \nonumber \\  
Z & = & \int d \nu \, det( M)^{n_f} \, 
          exp( \frac{\beta}{6} P ), \\
P & = & \sum_{(x_1, \ldots x_4)} Tr[ u(x_1, x_2) u(x_2, x_3) u(x_3, x_4)
u( x_4, x_1)]. \nonumber
\end{eqnarray}
Here $\beta$ is $6/g^2$ for bare gauge coupling constant $g$, $\nu$ is
the product of one copy of $SU(3)$ Haar measure for each link variable
on the lattice, and the sum in the definition of $P$ is over all nearest
neighbor squares $(x_1, \ldots x_4)$ with squares differing by a cyclic
permutation identified.  The extension of Eq.~(\ref{expG}) to vacuum
expectations of products of quark and antiquark fields and to QCD with
quarks having several different masses is not needed for the present
discussion and will be omitted for simplicity. In the present
discussion, $n_f$ can be either even or odd.

The valence approximation for $< G>$ is
\begin{eqnarray}
\label{expGv}
< G >_v & = & Z_v^{-1} \int d \nu \, G \,  
          exp( \frac{{\beta}_v}{6} P ), \nonumber \\  
Z_v & = & \int d \nu \, 
          exp( \frac{{\beta}_v}{6} P ). 
\end{eqnarray}
Here $\beta_v$ is $6/{g_v}^2$ with valence
approximation bare gauge coupling $g_v$. It is convenient to name the
shift between ${\beta}_v$ and $\beta$
\begin{eqnarray}
\label{defdel}
\Delta \beta & = & {\beta}_v - \beta.
\end{eqnarray}
The determination of $\Delta \beta$ will be discussed in
Section~\ref{sect:deltabeta}.  
As mentioned in Section~\ref{sect:intro},
$g_v$ and $g$ may also be viewed as related by a dielectric constant
$\eta$
\begin{eqnarray}
\label{defeta}
g_v & = & \frac{g}{\eta}.
\end{eqnarray} 
The calculation of $\Delta \beta$ in Section~\ref{sect:deltabeta} is, in
effect, also a calculation of $\eta$.  Although $\eta$ is useful in
describing the intuitive content of the valence approximation, it will
not appear directly in the remainder of this paper.

\section{ERROR EXPANSION}\label{sect:exp}

A coupling constant and quark loop expansion can now be constructed for
the difference between the full QCD vacuum expectation $< G>$ and its
valence approximation $< G>_v$.

It is convenient~\cite{Sexton} to express $< G>$ of Eq.~(\ref{expG}) as
\begin{eqnarray}
\label{expGr}
< G > & = & Z^{-1} \int d \nu \, G \,  
          exp( \frac{{\beta}_v }{6} P + Q) \nonumber \\
Z & = & \int d \nu \,  exp( \frac{{\beta}_v }{6} P + Q ), \\
Q & = & n_f tr {\log}(M) - \frac{\Delta \beta}{6} P. \nonumber
\end{eqnarray}
Introducing a parameter $\lambda$ multiplying $Q$, we expand $< G>$ in
powers of $\lambda$, replace $\lambda$ by 1, and get
\begin{eqnarray}
\label{errG}
< G > & = & < G>_v + \sum_n \delta_n (G),  \\ 
\label{delta1}
\delta_1( G) & = & < (G - <G>_v)(Q - <Q>_v)>_v, \\
\label{delta2}
\delta_2( G) & = & < ( G - < G>_v) (Q - < Q>_v)^2 >_v, \\
\label{delta3}
\delta_3( G) & = & 
< ( G - < G>_v) (Q - < Q>_v)^3 >_v - \nonumber \\  
& & 3 < ( G - < G>_v) (Q - < Q>_v) >_v < (Q - < Q>_v)^2 >_v, \\ 
\vdots \nonumber
\end{eqnarray}

In a coupling constant perturbation expansion of Eq.~\ref{errG} for the
difference between $< G>$ and $< G>_v$, the quantity $Q$ carries a
single quark loop. As a consequence $\delta_n (G)$ can be associated
with diagrams containing $n$ internal quark loops. None of the $\delta_n
(G)$, however, are simply sums of $n$-quark-loop diagrams . Each
includes also, through $Q$, counterterms arising from the shift between
$\beta$ of full QCD and the screened $\beta_v$ of the valence
approximation. We will return to this observation in Section~\ref{sect:mix}.

The quantity $tr {\log}(M)$ in $Q$ of Eq.~\ref{expGr} we now express as
a coupling constant power series. This series is also, formally, an
expansion in powers of a gauge potential. Thus, as usual in gauge field
theories, we use the theory's gauge invariance to transform to a gauge
which will tend to make the gauge potential small. For this purpose, we
transform the gauge field $u( x, y)$ to a euclidean lattice version of
Landau gauge.  The field $u( x, y)$ has been transformed to lattice
Landau gauge if for every lattice site $x$ the target function $\sum_y
tr[ u( x, y)]$ is a local maximum with respect to further gauge
transformations.  Generally there are many Gribov copies of
transformations taking a particular gauge field to euclidean lattice
Landua gauge.  The vacuum expectation value of any integrable function
of the gauge field transformed to Landau gauge is then an weighted
average over the Gribov copies of each field. The weights depend on the
particular choice of algorithm for obtaining Landau gauge.  The gauge
fixing algorithm used in our trial calculation in Section~\ref{sect:exp}
is discussed in Ref.~\cite{Sexton}.  Our expansion does not depend
explicitly on the choice of gauge fixing algorithm and, therefore, on
the choice of Gribov copy weighting.  We have not examined to what
degree this choice might be optimized to further minimize the gauge
potential and therefore speed the convergence of the coupling constant
expansion.

For each fixed gauge configuration, we construct a free quark coupling
matrix $M_0$ which approximates the interacting coupling matrix $M$ of
Eq.~\ref{defM}. For each configuration, let $z$ be the average over all
lattice links of $tr[ u( x, y)]/3$. Let $M_0$ be a free coupling matrix
with hopping constant $\kappa_0$ chosen to give a quark mass with agrees
with the mean-field-improved~\cite{Lepage} estimate
\begin{eqnarray}
\label{defM0} 
\frac{1}{2\kappa_0} - 4 & = & \frac{1}{2 z \kappa} - \frac{1}{2 z \kappa_c}, 
\end{eqnarray}
where $\kappa$ and $\kappa_c$ are, respectively, the hopping constant of
$M$ and the valence approximation to the critical value of this hopping
constant. The critical hopping constant is the smallest value for which
the pion mass becomes zero.  On the right side of Eq.~\ref{defM0}, the
parameter $z$ varies with gauge configuration but $\kappa$ and
$\kappa_c$ do not.

Mean-field improved perturbation theory suggests $z M_0$ as an
approximation to $M$.  We therefore express $tr {\log}(M)$ in the form
\begin{eqnarray}
\label{Mexp1}
tr {\log}(M) = tr {\log}\{ z M_0[ 1 - M_0^{-1}(M_0 - z^{-1}M)]\}, 
\end{eqnarray}
and expand to obtain the
\begin{eqnarray}
\label{Mexp2}
tr {\log}(M) = tr {\log}( z M_0) -  
\sum_n \frac{1}{n} tr\{[M_0^{-1}(M_0 - z^{-1} M)]^n\}. 
\end{eqnarray}
For small values of the chromoelectric potential, $M_0^{-1}(M_0 -
z^{-1}M)$ is linear in the potential. Thus Eq.~(\ref{Mexp2}) is
approximately an expansion in power of the chromoelectric potential or,
equivalently, in powers of a coupling constant.

To evaluate the trace in the second term of Eq.~(\ref{Mexp2}) we use an
ensemble of complex-valued quark fields $\phi$.  For each site $x$ and
each of the 12 combinations of spin index $s$ and color index $c$, the
corresponding $\phi_{sc}(x)$ we take to be an independent complex random
variable with absolute value 1 and probability distribution uniform on
the unit circle. For an $R$ element ensemble of such fields $\phi^r$, $1
\le r \le R$, we then have
\begin{eqnarray}
\label{Mexp3}
tr {\log}(M) = tr {\log}( z M_0) -  
R^{-1} \sum_{nr} \frac{1}{n} < \phi^r, [M_0^{-1}(M_0 -
z^{-1} M)]^n \phi^r>. 
\end{eqnarray}
Here $< \ldots , \ldots >$ is the inner product 
\begin{eqnarray}
\label{definner}
< \phi , \phi'> & = & \sum_{x s c} \phi^*_{sc}(x) \phi'_{sc}(x)
\end{eqnarray}
on the space of complex-valued quark fields.  In Eq.~(\ref{Mexp3}) the
quantity $tr {\log}( z M_0)$ and the inverse $M_0^{-1}$ can be both be
found easily in momentum space since $M_0$, in momentum space, is
block-diagonal.  To multiply vectors specified in position space by
$M_0^{-1}$, we transform in and out of momentum space by fast Fourier
transforms.

If the right side of Eq.~(\ref{Mexp3}) is substituted for $tr {\log}(M)$
in the expression for $Q$ in Eq.~\ref{expGr}, Eq.~\ref{errG} becomes a
coupling constant and quark-loop expansion for corrections to valence
approximation vacuum expectation values.  The only quantity in this
expansion not yet specified is the shift $\Delta \beta$.
Eqs.~(\ref{errG}) and (\ref{Mexp3}) are formally correct for any choice
of $\Delta \beta$. The rate at which these series converge, however,
will be affected by this choice.

\section{$\Delta \beta$}\label{sect:deltabeta}

For valence approximation calculations of the light hadron spectrum,
the up and down quark masses are usually taken to be equal and the
corresponding $\kappa$ is chosen by requiring the pion mass to have its
physical value. The valence approximation $\beta_v$ is then determined by
setting the rho mass to its physical value. Thus, in effect,
$\Delta \beta$ is found by requiring the valence approximation
error in the rho mass to vanish. Since the rho mass is expected to be
determined mainly by the low-momentum behavior of the chromoelectric
field, this choice may be viewed as a quantitative implementation of the
qualitative picture of the valence approximation mentioned briefly in
Sect.~\ref{sect:intro}.  A class of possible alternatives consists of
choices of $\Delta \beta$ which make the error, or some approximation to
the error, equal to zero for other quantities beside the rho mass
which are determined mainly by the low-momentum behavior of the
chromoelectric field.  A convenient version of this idea for the present
discussion is to fix $\Delta \beta$ by requiring zero valence
approximation error for the Landau gauge gluon propagator at minimal
nonzero momentum to first order in quark loops and to second order in
the coupling constant expansion.  As an additional simplification, the
expectation values needed to determine $\Delta \beta$ we find using
(analytic) mean-field-improved perturbation theory rather than by Monte
Carlo. For the test case considered in Section~\ref{sect:exp}, we obtain
the same value of $\Delta \beta$ by this method as produced by the
non-perturbative method of Ref.~\cite{Sexton}.

The requirement for $\Delta \beta$ becomes that $\delta_1(G)$ of
Eq.~\ref{delta1} vanishes
\begin{eqnarray}
\label{defdeltabeta}
< (G - <G>_v)(Q - <Q>_v)>_v & = & 0, 
\end{eqnarray}
where
\begin{eqnarray}
G & = & \sum_{c\mu} \tilde{A}_{c\mu}(p) \tilde{A}_{c\mu}(-p), \\
Q & = & -tr( V) - \frac{1}{2} tr( V^2) - \frac{\Delta \beta}{6} P, \nonumber \\
V & = & M_0^{-1}(M_0 - z^{-1} M). \nonumber
\end{eqnarray}
The quantity $\tilde{A}_{c\mu}(p)$ is the Fourier transformed gauge potential
\begin{eqnarray}
\tilde{A}_{c\mu}(p) & = & \sum_x exp(-ip \cdot x) A_{c\mu}( x).
\end{eqnarray} 
for momentum vector $p$ with a single nonzero component $p_1$ of
$2 \pi/L$, where $L$ is the lattice period, 
To second order in mean-field-improved perturbation theory, 
the link field $u(x,y)$, in $M$, and the plaquette action $P$ can be
approximated by
\begin{eqnarray}
\label{mfipuP}
u( x, x + \hat{\mu}) & = & z \{1 -i A_{c \mu}( x + \frac{1}{2}\hat{\mu}) T_c - 
\frac{1}{2} [A_{c \mu}( x + \frac{1}{2}\hat{\mu}) T_c]^2\}
\nonumber \\ 
P & = & \frac{<tr U >_v}{3}\sum_{x c \mu \nu}[ 1 - \frac{1}{2}( F_{c\mu\nu}( x +
\frac{1}{2}\hat{\mu} + \frac{1}{2}\hat{\nu})^2] \\
F_{c\mu\nu}( x) & = & A_{c\mu}( x + \frac{1}{2}\hat{\nu}) - 
A_{c\mu}( x - \frac{1}{2}\hat{\nu}) +
A_{c\nu}( x - \frac{1}{2}\hat{\mu}) - A_{c\nu}( x + \frac{1}{2}\hat{\mu}). \nonumber
\end{eqnarray}
where the $T_c$ are an orthnormal basis for the Lie algebra of $SU(3)$ 
\begin{eqnarray}
tr( T_c T_d) & = & \frac{1}{2} \delta_{cd}, \nonumber
\end{eqnarray}
and $<tr U>_v$ is the valence approximation plaquette expectation value.

\begin{figure}                                                   
\epsfxsize=\textwidth                                            
\epsfbox{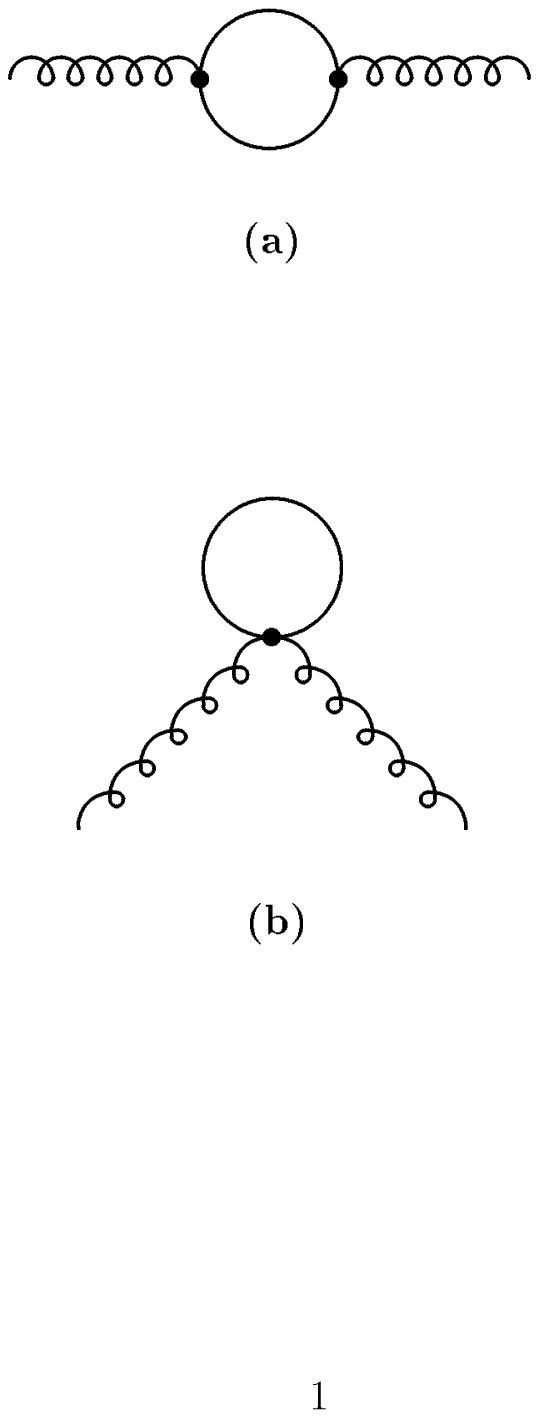}                                             
\caption{Feynman diagrams which contribute to $\Delta \beta$.}   
\label{fig:vac_pol}                                              
\end{figure}

The vacuum expectation values in Eqs.\ref{deltabeta} we evaluate by
lattice weak coupling perturbation theory to second order in the valence
approximation coupling constant. This calculation reduces to finding the
two vacuum polarization Feynman diagrams in Fig.~\ref{fig:vac_pol}.
These diagrams for QCD are proportional to the corresponding diagrams
for a $U(1)$ lattice gauge theory and thus fulfill the $U(1)$ theory's
Ward identities. The calculation of the tadpole diagram,
Figure~\ref{fig:vac_pol} (b) can thus be eliminated.
We obtain
\begin{eqnarray}
\label{deltabeta}
\Delta \beta  & = &  
\frac{9 n_f}{4 \sin^2(\pi/L) <tr U >_v} [\Pi_{2 2}(p) - \Pi_{2 2}(0)] \\
\Pi_{\mu \nu}(p) & = &
\frac{1}{L^4} \sum_q tr[ \Gamma_{\mu}(q + p/2 ) S(q + p) 
\Gamma_{\nu}(q + p/2 ) S(q) ] \nonumber,
\end{eqnarray}
where each component of $q$ in the sum over $q$ ranges from $\pi/L$ to 
$2\pi - \pi/L$ in steps of $2\pi/L$. The propagator 
$S(q)$ and vertex $\Gamma_{\mu}(q)$ are
\begin{eqnarray}
\label{sandgamma}
S(q) & = & \frac{1}{1/(2 \kappa_0) - i \sum_{\mu} \gamma_{\mu} sin( q_{\mu}) - 
\sum_{\mu} cos( q_{\mu})}, \nonumber \\
\Gamma_{\mu}(q) & = & sin( q_{\mu}) - i\gamma_{\mu} cos(q_{\mu}). 
\end{eqnarray}

The limiting value of $\Delta \beta$ for large $L$ without mean-field
improvement has been derived in Ref.~\cite{Hasenfratz}.

\section{EXAMPLE}\label{sect:examp}

As a test of our method we compared valence approximation expectations
$< G>_v$, Eq.~(\ref{expGv}), their one-loop errors $< (G - <G>_v)(Q -
<Q>_v)>_v$, Eq.~(\ref{errG}), and the corresponding full QCD
expectations $<G>$, Eq.~(\ref{expGr}), for a lattice $10^4$ with
$\beta_v$ of 5.679, $\kappa$ of 0.16 and $n_f$ of 2. For a $16^3 \times
32$ lattice at $\beta_v$ of 5.70, Ref~\cite{Butler_mass} gives a
critical $\kappa_c$ of 0.16940(5) and strange quark mass $m_s a$ in
lattice units of 0.097(3). Thus $\kappa$ of 0.16 corresponds to a quark
mass about 1.8 times $m_s a$.  According to Eq.~(\ref{deltabeta}),
$\Delta \beta$ is 0.243 giving a full QCD $\beta$ of 5.436. For this case
$\Delta \beta$ found by the method of Ref.~\cite{Sexton} is 0.244(6).

We used 224 independent equilibrium gauge configuration in the valence
approximation ensemble, generated by an over-relaxed pseudo heat bath
algorithm, 600 random fermion fields $R$ to evaluate the trace in
Eq.~(\ref{Mexp3}) and 176 weakly correlated equilibrium gauge configurations
for the full QCD ensemble, generated by a red-black preconditioned
hybrid Monte Carlo algorithm. The expansion in Eq.~(\ref{Mexp3}) was
carried to order $n$ of 10.
The calculation of $< (G - <G>_v)(Q - <Q>_v)>_v$  was not
turned carefully. In particular $R$ of 600 in Eq.~(\ref{Mexp3}) is much larger than its
optimal vaule. The time required for the valence approximation and error
calculation was still less than 5\% of the time required by the full QCD
calculations.

\begin{figure}
\epsfxsize=\textwidth
\epsfbox{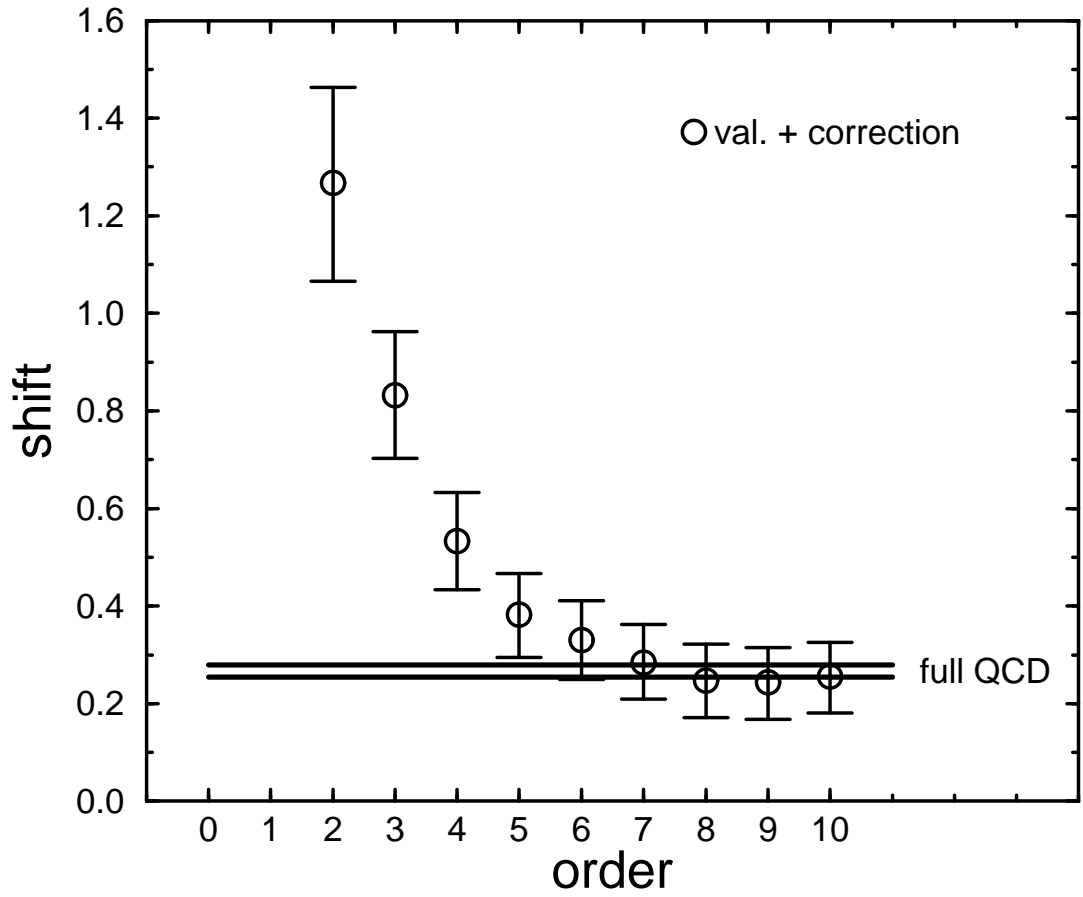}
\caption{
The predicted relative shift in Wilson loop $W_{10}$ from its valence
approximation value as a function of the order $n$ in coupling constant
compared to the true shift of full QCD.}
\label{fig:loop10}  
\end{figure}

\begin{figure}
\epsfxsize=\textwidth  
\epsfbox{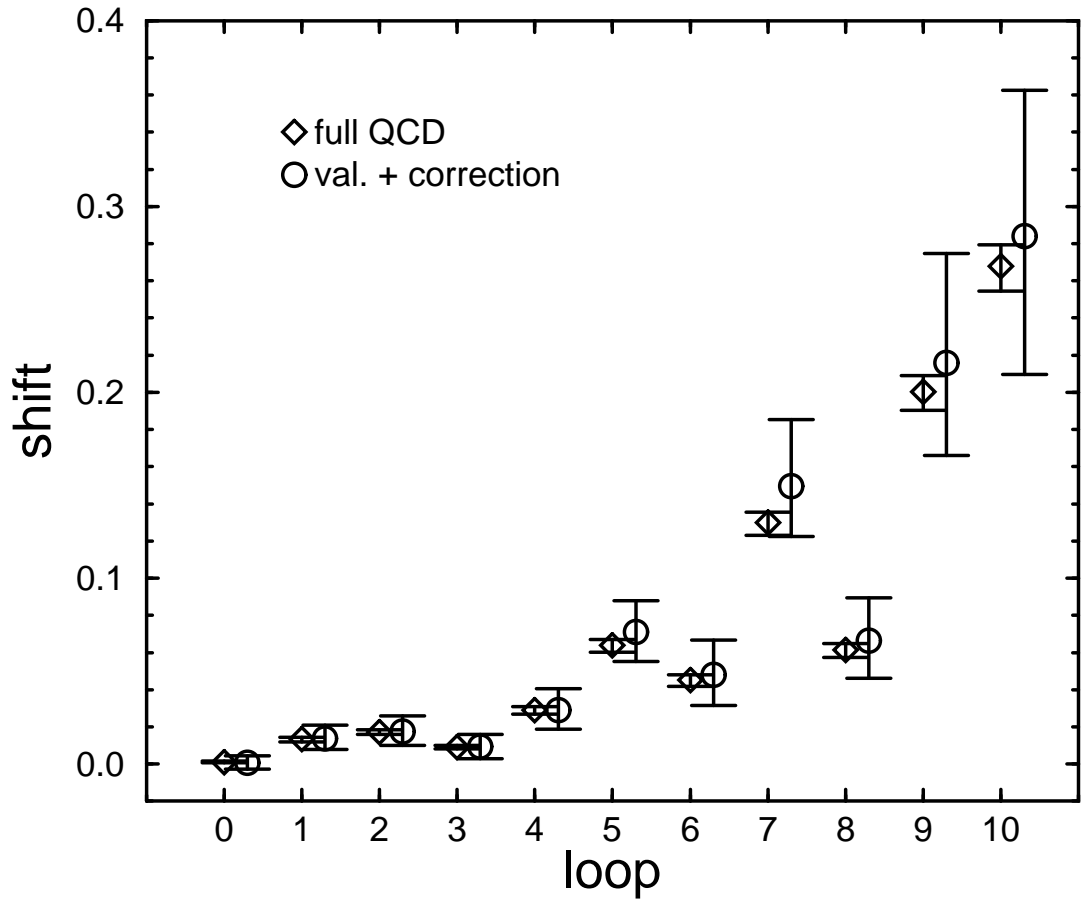}
\caption{The predicted relative shift in 11 Wilson loops from their
valence approximation values in comparison to the true shift of full QCD.}
\label{fig:order7}  
\end{figure}

For $G$ we used Wilson loops $W_0, \ldots W_{10}$ consisting,
respectively, of paths $1 \times 1$, $2 \times 1$, all rotations of
steps in the directions $\hat{1},\hat{2},\hat{3},-\hat{1},-\hat{2},
-\hat{3}$, all rotations of steps in the directions
$\hat{1},\hat{2},\hat{3},-\hat{2},-\hat{1}, -\hat{3}$, $3 \times 1$, $2
\times 2$, $4 \times 1$, $5 \times 1$, $3 \times 2$, $4
\times 2$, and $3 \times 3$. 
For the $3 \times 3$ loop, $W_{10}$, Fig.~\ref{fig:loop10} shows the
predicted error $< (W_{10} - <W_{10}>_v)(Q - <Q>_v)>_v$ as a function of
the highest power $n$ of coupling strength used in Eq.~(\ref{Mexp3}).
The error converges adequately by $n$ of 7. For smaller Wilson loops,
$W_0,\ldots W_9$, the predicted error's convergence as a function of $n$
is comparable to or faster than the convergence shown in
Fig.~\ref{fig:loop10}.  For $n$ of 7, Fig.~\ref{fig:order7} shows the
relative shift of the valence approximation from full QCD $(<W_i> -
<W_i>_v)/<W_i>$ and the predicted value $< (W_i - <W_i>_v)(Q -
<Q>_v)>_v/<W_i>$.  To within statistical uncertainties, the predicted
errors agree with the true errors.

The true errors in Fig.~\ref{fig:order7} were found from the shortest
full QCD run sufficient to confirm equilibration of $<W_0>, \ldots
<W_{10}>$. Nonetheless the statistical uncertainties in the predicted
errors are much larger than those in the true errors.  If we were to run
the error prediction algorithm long enough to obtain statistical
uncertainties comparable to the uncertainties found by a direct comparison
of full QCD and the valence approximation, it is possible that the
computer time required by the error algorithm would become comparable to
that for full QCD. To find the uncertainty arising from use of the
valence approximation, however, the statistical uncertainty in the error
estimate does not need to be too much smaller than the error estimate's
central value. Used in this way, for the set of parameters of the test,
our algorithm takes significantly less time than the shortest possible
direct comparison of the valence approximation and full QCD.

\section{GLUEBALL-QUARKONIUM MIXING} \label{sect:mix}

We now consider briefly the valence approximation to glueball-quarkonium
mixing, corrections to the valence approximation to mixing which follow
from Eq.~(\ref{errG}) and a mixing calculation reported in
Ref.~(\cite{Boglione}).  The lowest lying glueball, according to the
valence approximation, is stable and is expected to be a scalar.
Evidence that $f_0(1710)$ is composed mainly of this state is given in
Refs.~\cite{Sexton_glue,Weingarten,Lee98a,Lee98b}.  With quark-antiquark
annihilation initially ignored, the lightest scalar quarkonium states
are also stable. Their valence approximation masses and evidence for
their identification with observed states are discussed in
Refs.~\cite{Lee97,Weingarten,Lee98a,Lee98b}.

\begin{figure}
\epsfxsize=\textwidth
\epsfbox{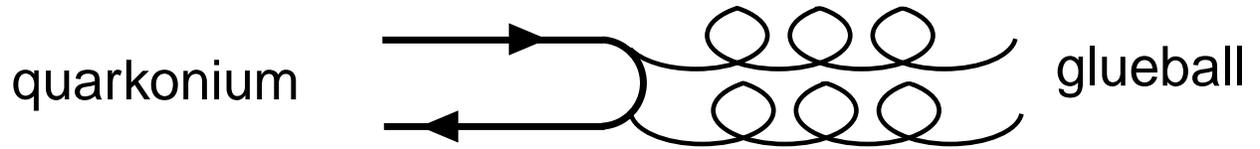}
\caption{Quarkonium-glueball mixing through quark-antiquark annihilation.}
\label{fig:mix}
\end{figure}

Mixing among the $s\overline{s}$ and $(u\overline{u} +
d\overline{d})/\sqrt{2}$ scalars and the scalar glueball then occurs
through quark-antiquark annihilation. In the valence approximation, the
glueball-quarkonium mixing energy can be extracted~\cite{Lee98a,Lee98b}
from the vacuum expectation value
\begin{eqnarray}
\label{gs}
C_v( t) & = & < g(t) s(0)>_v,
\end{eqnarray}
where $g(\vec{x},t)$ is the smeared zero-momentum scalar glueball
operator of Ref.~\cite{Vaccarino}, with vacuum expectation subtracted,
and $s(\vec{x},t)$ is the smeared zero-momentum scalar quarkonium
operator of Ref.~\cite{Lee97}.  It is convenient to define also a full
QCD $C(t)$ by Eq.~(\ref{gs}) with $< \ldots>_v$ replaced by $< \ldots
>$.  A qualitative representation of $C_v(t)$ is provided by
Figure~\ref{fig:mix} giving a typical Feynman diagram contributing to
the lattice weak coupling expansion for $C_v(t)$.

\begin{figure}
\epsfxsize=\textwidth
\epsfbox{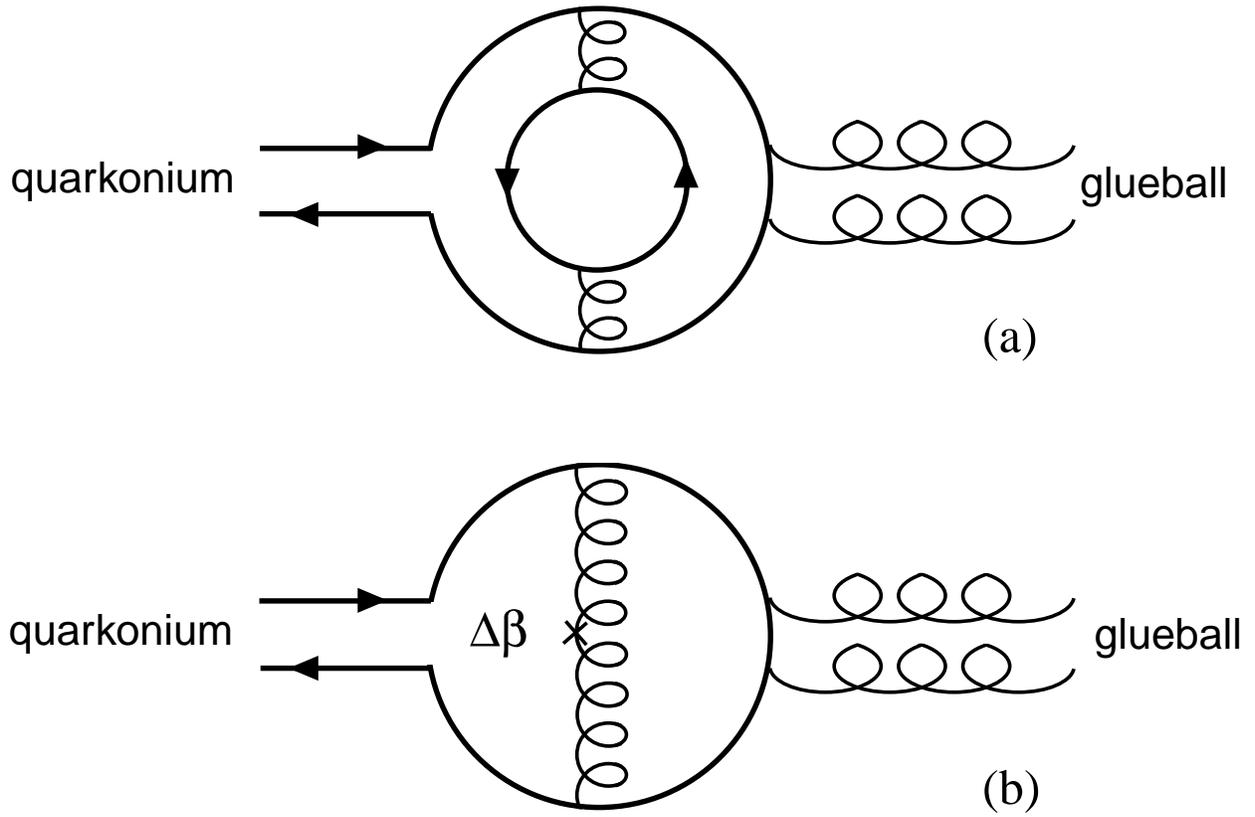}
\caption{One-quark-loop corrections to the valence approximation to
quarkonium-glueball mixing.}
\label{fig:mixcor}
\end{figure}

Assuming, for simplicity, only vacuum polarization arising from $u$
and $d$ quarks taken to have degenerate mass,
the one-quark-loop correction to $C_v(t)$ can be found from
Eqs.~(\ref{expGr}-\ref{delta1}). For the present
discussion, we will not apply the expansion of Eq.~(\ref{Mexp2}).
The one-quark-loop error in $C_v(t)$ becomes
\begin{eqnarray}
\label{delta1gs}
C(t) - C_v(t) & = & T_M - \frac{\Delta \beta}{6} T_P, \nonumber \\
T_M & = & < [g(t)s(0) - <g(t)s(0)>_v] [ 2 tr {\log}(M) - < 2 tr
{\log}(M)>_v] >_v \\
T_P & = & < [g(t)s(0) - <g(t)s(0)>_v] [ P - < P >_v]>_v. \nonumber
\end{eqnarray}  
Qualitative representations of $T_M$ and $T_P$ are given by typical
Feynman diagrams contributing to their weak coupling expansions shown in
Figure~\ref{fig:mixcor}(a) and Figure~\ref{fig:mixcor}(b), respectively.
Among the processes contributing to $T_M$ in Figure~\ref{fig:mixcor}(a)
are glueball-quarkonium transitions through common pi-pi, kaon-antikaon,
and eta-eta decay channels.  The quantity $T_P$, on the other hand, is
the counterterm, discussed in Section~\ref{sect:exp}, which arises from
the shift $\Delta \beta$ between $\beta$ of full QCD and the screened
$\beta_v$ of the valence approximation.

In Ref.~\cite{Boglione} a model is proposed for mixing among the valence
approximation to the lightest scalar glueball state and the valence
approximations to the lightest scalar $s\overline{s}$ and
$(u\overline{u} + d\overline{d})/\sqrt{2})$ states. Applied to
glueball-quarkonium mixing energies, this model omits the leading
valence approximation mixing amplitude coming from $C_v(t)$ and
represented in Figure~\ref{fig:mix}. The model includes instead only
transitions through common pi-pi, kaon-antikaon and eta-eta decay
channels. These transition do contribute to $T_M$. Thus the model might
be viewed as a calculation of quark-loop corrections to the valence
approximation to mixing if not as an evaluation of the full mixing
process.  The equation assumed to govern mixing between the valence
approximation glueball and quarkonium states through intermediate decay
channels, however, entirely ignores the counter-term $T_P$.  No argument
is offered in support of this omission.  Ref.~\cite{Boglione} simply
assumes, without proof, a relation between full QCD and valence
approximation propagators with no term corresponding to $T_P$.

With this counter-term dropped, Eq.~(\ref{delta1gs}) gives the error in
$C_v(t)$ for a version of the valence approximation with $\Delta \beta$
forced to zero.  Equivalently, it is easily checked that $T_P$ is the
derivative of $C_v(t)$ with respect to $\beta_v$. Thus by dropping the
$T_P$ term from Eq.~(\ref{delta1gs}) the one-quark-loop error estimate
for $C_v(t)$ is altered by approximately the increment in $C_v(t)$ in
going from $\beta_v - \Delta \beta$ to $\beta_v$ .  For $\beta_v$ of
5.93, $\Delta \beta$ is known to be greater than 0.23. Thus a lower
bound on the effect of setting $\Delta \beta$ to zero can be found by
comparing valence results at $\beta_v$ of 5.70 with those at $\beta_v$
of 5.93. The data in Refs.~\cite{Lee98a,Lee98b} then shows that the
one-quark-loop error estimate for glueball-quarkonium energy is changed
by an amount equal to the entire leading valence approximation to the
mixing energy obtained from $C_v(t)$.

A cross check on the consequences for valence approximation errors of
forcing $\Delta \beta$ to zero can be obtained by making this change in
the error formula applied to low-lying hadron masses and meson decay
constants.  Using the data in Refs.~\cite{Butler_mass,Butler_decay} for
$\beta_v$ of 5.70 and 5.93, we find that with $\Delta \beta$ forced to
zero masses and decay constants are off by as much as 45\%, rather than
by less than 10\% or less than 20\%, respectively, for an optimally
chosen $\Delta \beta$.

Thus as a calculation of errors in the valence approximation to
glueball-quarkonium mixing energies, Ref.~\cite{Boglione} would be
expected to predict significantly larger errors than actually occur with
an optimal choice of $\Delta \beta$.  As we mentioned earlier, however,
also missing from the calculation of Ref.~\cite{Boglione} is the leading
valence approximation term which can be obtained from $C_v( t)$.  
It appears to us that Ref.~\cite{Boglione} gives neither an adequate
model of the full glueball-quarkonium mixing process nor of the
corrections to the leading valence approximation to this process. We
believe its results are simply incorrect.

In partial defense of Ref.~\cite{Boglione}, it might be argued that
although $T_P$ is not explicitly present in the relation given between
valence approximation propagators and those of full QCD, $T_P$ is
nonetheless present implicitly. The coupling between valence
approximation states and two-body decay channels is assumed to fall
exponentially with $|\vec{k}|^2$, where $\vec{k}$ is the center-of-mass
system 3-momentum carried by one of the decay products.  Perhaps this
exponential cutoff removes from the equation of Ref.~\cite{Boglione}
those contributions which $T_P$ subtracts from our equations. For this
to hold would require a surprising coincidence since no mention is made
in Ref.~\cite{Boglione} of the need for a term like $T_P$ in the
relation between full QCD and the valence approximation and no attempt
is made to tune the cutoff to absorb this term.  The cutoff is
introduced simply as the authors' expectation of the behavior of
coupling between unstable scalars and their pseudoscalar decay products.

In addition, however, it is mentioned explicitly in
Ref.~\cite{Boglione}, and supported by the tables giving proposed values
of full QCD corrections to valence approximation masses, that the model
of Ref.~\cite{Boglione} predicts full QCD masses below valence
approximation masses for those states which are stable in the valence
approximation but unstable in full QCD.  This is exactly the result to
be expected for corrections to the valence approximation given by
Eqs.~(\ref{errG}) and (\ref{delta1}) with $T_P$ removed. For the vacuum
expectation of an arbitrary $G$, Eqs.~(\ref{errG}) and (\ref{delta1})
give
\begin{eqnarray}
\label{delta1G}
<G> - <G>_v & = & T^G_M - \frac{\Delta \beta}{6} T^G_P,\nonumber \\
T^G_M & = & < (G - < G >_v) [ 2 tr {\log}(M) - < 2 tr {\log}(M)>_v] >_v \\
T^G_P & = & < (G - < G >_v) ( P - < P >_v)>_v. \nonumber
\end{eqnarray}
The contribution to the error in $<G>_v$ from the term $T^G_M$ is the
incremental effect of the color charge screening due to a single quark
loop and therefore of a decrease, by some amount, in the QCD effective
charge.  As a consequence of QCD's asymptotic freedom this term shifts
quantities with mass units toward smaller values in full QCD than in the
valence approximation.  From our discussion earlier it follows that the
term $T^G_P$ has the opposite effect. It shifts quantities with mass
units toward larger values in full QCD than in the valence
approximation.  In fact, as might be expected from the discussion of
Ref.~\cite{Weingarten}, calculations of valence approximation decay
constants \cite{Butler_decay,CPPACS} and a recent calculation of masses
\cite{CPPACS} show that full QCD quantities for excited states are
consistently larger those of the valence approximation.  Therefore $T^G_P$ for
propagators of excited states is consistently larger in magnitude than
$T^G_M$, and the model of Ref.~\cite{Boglione} predicts even the wrong
sign for the relation between masses in full QCD and in the valence
approximation. It appears to us this error is clear evidence that the
model's cutoff on decay momenta can not have absorbed the effect of the
omission of $T^G_P$.

\newpage

\end{document}